\documentclass[journal]{IEEEtran}
\usepackage{siunitx}
\usepackage{siunitx}
\usepackage{graphicx}
\usepackage{algorithm,algpseudocode}
\usepackage[utf8]{inputenc}
\usepackage{cite}
\usepackage{float}

\begin{document}
\title{Terahertz dynamic aperture imaging at stand-off distances using a Compressed Sensing protocol}
\author{{S.~Augustin, P.~Jung, S. Frohmann, H.-W.~H\"ubers}
\thanks{S. Augustin and H.-W. H\"ubers are with the Department of Physics Humboldt-Universit\"at zu Berlin \& the Institute of Optical Sensor Systems of the German Aerospace Center Berlin.}
\thanks{P. Jung is with the Department of Electrical Engineering and Computer Science, Technische Universit\"at Berlin.}
\thanks{S. Frohmann is with the Institute of Optical Sensor Systems of the German Aerospace Center Berlin.}
\thanks{Manuscript received ...; revised ....}}
\maketitle
\begin{abstract}
In this text, results of a \SI{0.35}{} terahertz (THz) dynamic aperture imaging approach are presented. The experiments use an optical modulation approach and a single pixel detector at a stand-off imaging distance of $\approx$\SI{1}{\meter}. The optical modulation creates dynamic apertures  of \SI{5}{\centi\meter} diameter with $\approx$2000 individually controllable elements. An optical modulation approach is used here for the first time at a large far-field distance, for the investigation of various test targets in a field-of-view of 8 x \SI{8}{\centi\meter}. The results highlight the versatility of this modulation technique and show that this imaging paradigm is applicable even at large far-field distances. It proves the feasibility of this imaging approach for potential applications like stand-off security imaging or far field THz microscopy.
\end{abstract}
%
%
\section{Introduction}
\IEEEPARstart{T}{erahertz} (THz) imaging is a powerful technique with many applications ranging from nondestructive testing \cite{Amenabar.2013} and biomedicine \cite{Smye.2001} to security \cite{Yurduseven.2014}. For security applications stand-off imaging, i.e. the detection and possibly the identification of hidden objects at distances of several meters, is of particular interest. Therefore, a number of THz stand-off imaging systems have been developed by several research groups. These are either active systems, relying on artificial THz illumination with a narrow-band THz source \cite{Cooper2014} or passive systems, which are in some sense the THz analogue to infrared cameras since these systems detect the natural THz radiation emitted or reflected from a person or object \cite{Heinz.2015}. Both imaging approaches have a low number of detectors - typically one for active imagers and up to about 100 for passive imagers. Therefore, additional scanning of the scene is required. It is usually realized by mechanical scanning of a mirror. Such an approach has several shortcomings, for example a limited frame rate, imaging artifacts, and limited long-term reliability. An alternative imaging approach that facilitates so-called digital beam forming \cite{Ahmed.2011}, uses a sparse array of transmitters (TXs) and receivers (RXs) for close-by imaging. While this approach is appealing since it involves no mechanical scanning, it also is very demanding when it comes to the necessary hardware, in particular the large number of TXs and RXs.
\vspace{3pt}\\ \indent
A related approach makes use of a single TX in combination with a single detector pixel. Such a single-pixel THz imaging system has been demonstrated in 2008 \cite{Chan.2008}. In this case, spatially-structured illumination is generated by a set of patterns (masks) and the image is formed without mechanical scanning using a Compressed Sensing (CS) algorithm. However, due to the lack of spatial light modulators  (SLMs), which are needed for dynamically changing the patterns, the patterns had to be changed by hand and, therefore, this approach initially did not find widespread use in THz imaging. This changed in recent years due to the availability of SLMs. Following the initial idea various configurations have been realized since then \cite{Busch.2012, Shrekenhamer.2013, Augustin.2015, Xie.2013}. In these systems THz sources such as a THz time-domain system \cite{Busch.2012},\cite{Xie.2013}, a multiplier source \cite{Augustin.2015} or an incandescent light source \cite{Shrekenhamer.2013} have been used in combination with a single detector pixel. To achieve the dynamic change of the  patterns an optical modulation technique is combined with a SLM for visible light. This technique allows single-pixel imaging functionality without mechanical scanning during the automatic image acquisition. The visible light (VIS) SLM is used to project spatial patterns (masks) onto a suitable semiconductor disc, the optical switch (OS). In the illuminated regions electrons are excited into the conduction band and the semiconductor partially changes from a semiconducting to a metallic state, becoming less transmissive for THz radiation. The performance of such an optically controllable single detector pixel camera is influenced by several factors including the quality of the OS’s illumination and the semiconductor material itself \cite{Gallacher.2012}, \cite{Kannegulla.2015}. A THz image of the scene is formed from sequential measurements of the THz responses when different VIS-masks are commanded to the SLM.
In the image reconstruction/decoding process of the sequential measurements a system of linear equations, which can even be underdetermined, is solved using for example Compressed Sensing algorithms. Although, the optical modulation approach has already proven to be very versatile and powerful, so far this imaging approach has only been demonstrated for object distances of a few millimeters to a few centimeters or in a near-field THz configuration \cite{Watts.2014, Augustin.2018, Mohr.2018, Stantchev.2017, Stantchev.2016}. This potentially limits the applicability of this imaging approach.  
\vspace{3pt}\\ \indent
In some of the most promising applications of THz imaging, in particular in security imaging, it is mandatory to obtain imagery at stand-off distances. Due to the lack of sensitive, multi-pixel THz cameras, a single pixel camera without mechanical scanning is a promising solution. However, whether dynamic aperture imaging with a SLM and a single pixel detector can be implemented in a stand-off modality is still an open question. The special challenge in this comes from the large distance between the OS and the scene. The details of this special challenge are discussed in the next section. Nonetheless, it is the objective of this work to evaluate the potential of dynamic aperture imaging in combination with Compressed Sensing (CS) for stand-off imaging at THz frequencies. 
\section{Special Challenge at stand-off distances}\label{sec:challenge}
\begin{figure}[H]
\centering
\includegraphics[width=\linewidth]{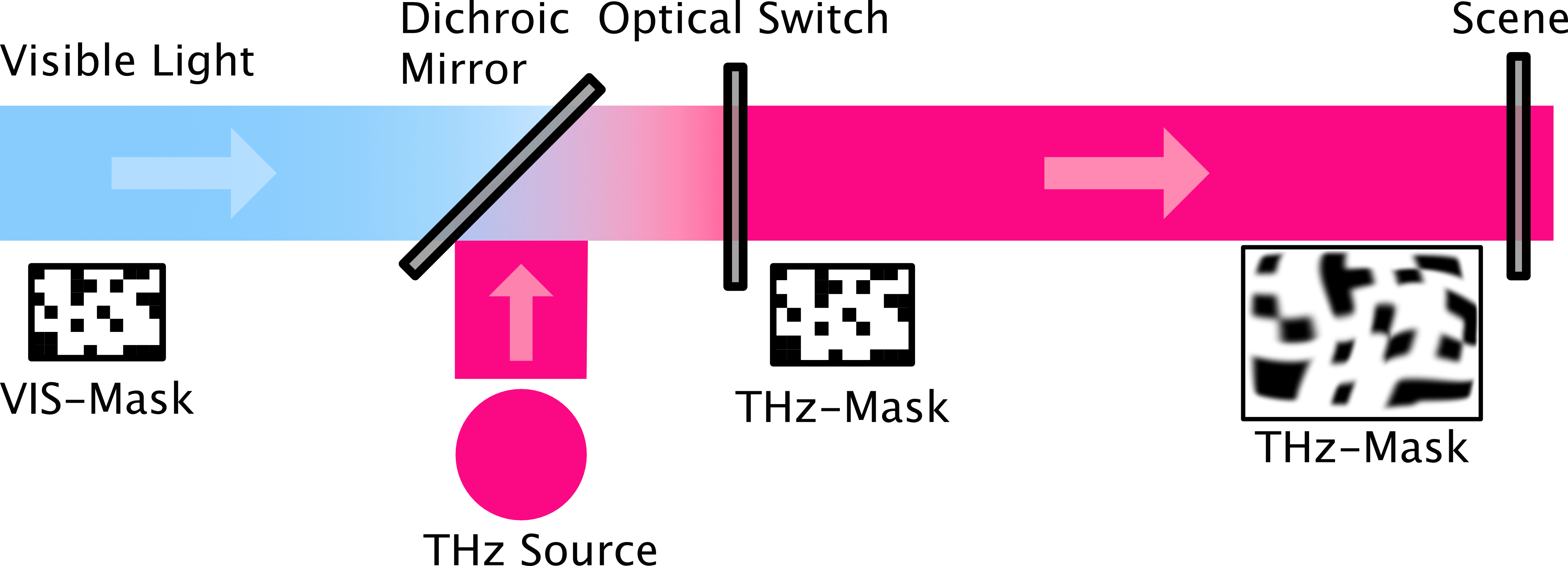}
\caption{Visualization of the challenge of the optical modulation approach illustrating the THz-mask degradation over the distance from the OS to the scene.}
\label{fig:distortion}
\end{figure}

\begin{figure}[H]
\centering
\includegraphics[width=\linewidth]{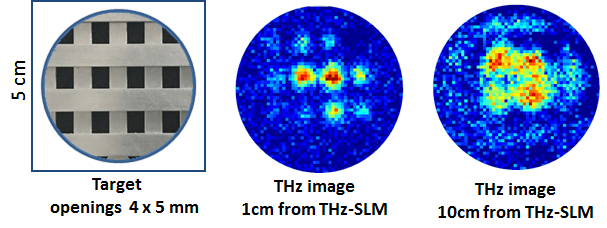}
\caption{THz transmission images of a metal grid at \SI{1}{\centi\meter} and \SI{10}{\centi\meter} distance from the OS, illustrating the influence of mask degradation on the image quality.}
\label{fig:masks}
\end{figure}

Some of the main advantages of dynamic aperture imaging, i.e. the achievable spatial resolution and the signal-to-noise ratio (SNR) as well as the overall fidelity of the image acquisition depend crucially on the fidelity of the THz-masks at the position of the scene. Ideally, the THz-masks are identical to the commanded VIS-masks. However, due to beam divergence and diffraction, the THz-masks are somewhat distorted. This special challenge of the optical modulation approach is visualized in Fig. \ref{fig:distortion}. The figure shows that the VIS-masks, which are commanded to the VIS-SLM are equal to the resulting THz-masks only at the surface of the optical switch (OS). Since in a stand-off modality the scene is a large distance away from the OS, the THz-masks incur a deviation from their initial state (beam divergence, diffraction, etc.). This deviation from the commanded masks increases with the distance from the OS to the scene and limits the achievable spatial resolution \cite{Stantchev.2016}.
As discussed, it is obvious that this effect is especially challenging in a stand-off modality. Therefore, we have first analyzed the severity of the THz-mask degradation. This was done in a separate single detector pixel imaging, transmission configuration (see \cite{Augustin.2016} for setup details). This transmission setup was used to image a metal grid that was placed at two distances (\SI{1}{\centi\meter} and \SI{10}{\centi\meter}) behind the output aperture of the THz-SLM. The radiation from a \SI{0.35}{\tera\hertz} source, which is transmitted through the grid was focused onto a heterodyne detector and the image was reconstructed from 30.000 masks using a Compressed Sensing algorithm. The result is shown in Fig. \ref{fig:masks}. The metal test target is clearly discernible at \SI{1}{\centi\meter} distance from the OS. When the distance is increased to \SI{10}{\centi\meter} the shape of the grid becomes blurred due to divergence and diffraction of THz beam. This indirectly demonstrates that the THz-masks severely degrade already with an increasing distance from the OS on the centimeter scale. This is obviously a major challenge for stand-off imaging, where the scene is several meters away from the THz-SLM.
%
%
\section{Methods}
\subsection{Optical Modulation Approach and Experimental Setup}
\begin{figure}[H]
\centering
\includegraphics[width=\linewidth]{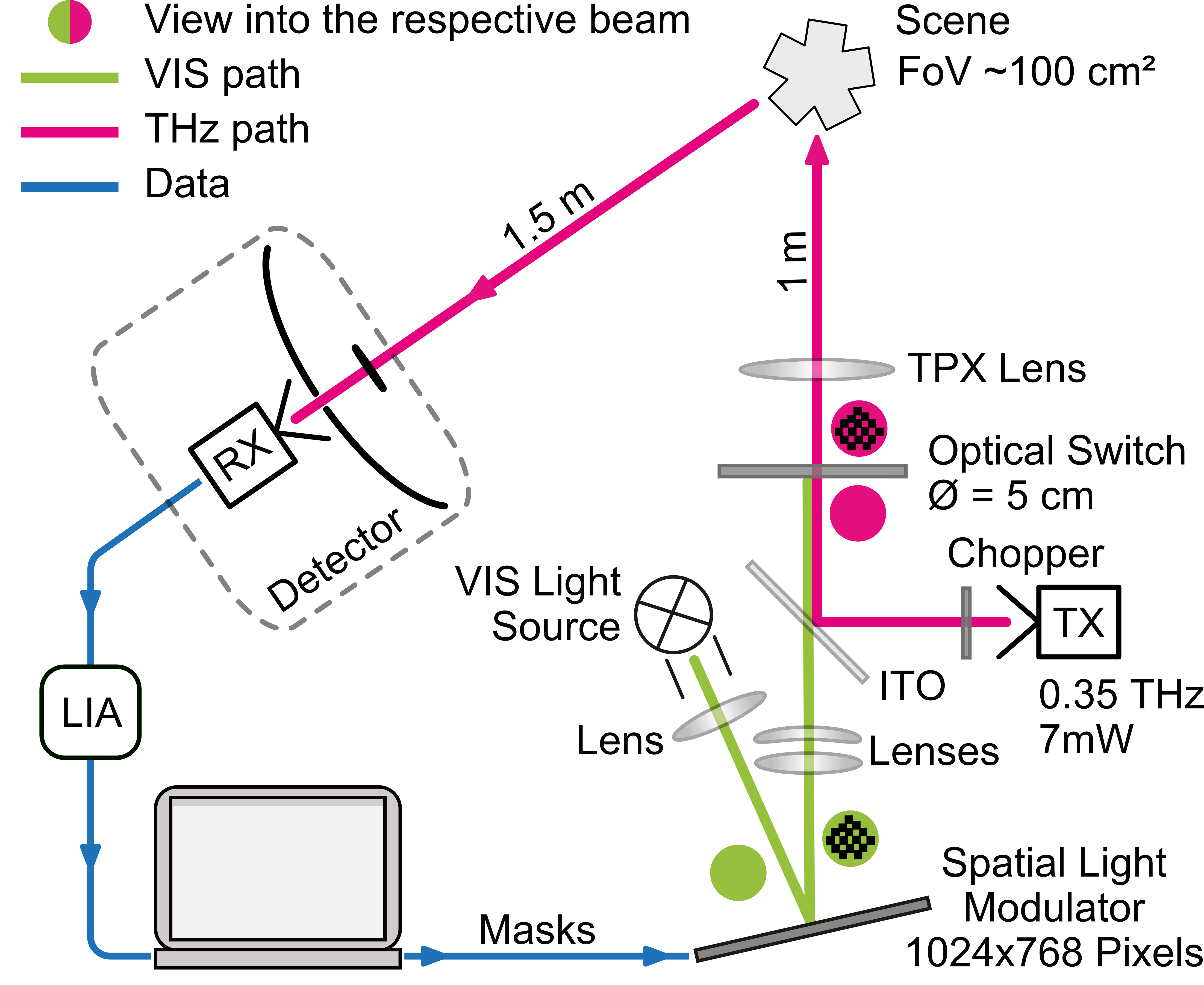}
\caption{Block Diagram of the experimental setup.}
\label{fig:setup}
\end{figure}
\begin{figure}[H]
\centering
\includegraphics[width=\linewidth]{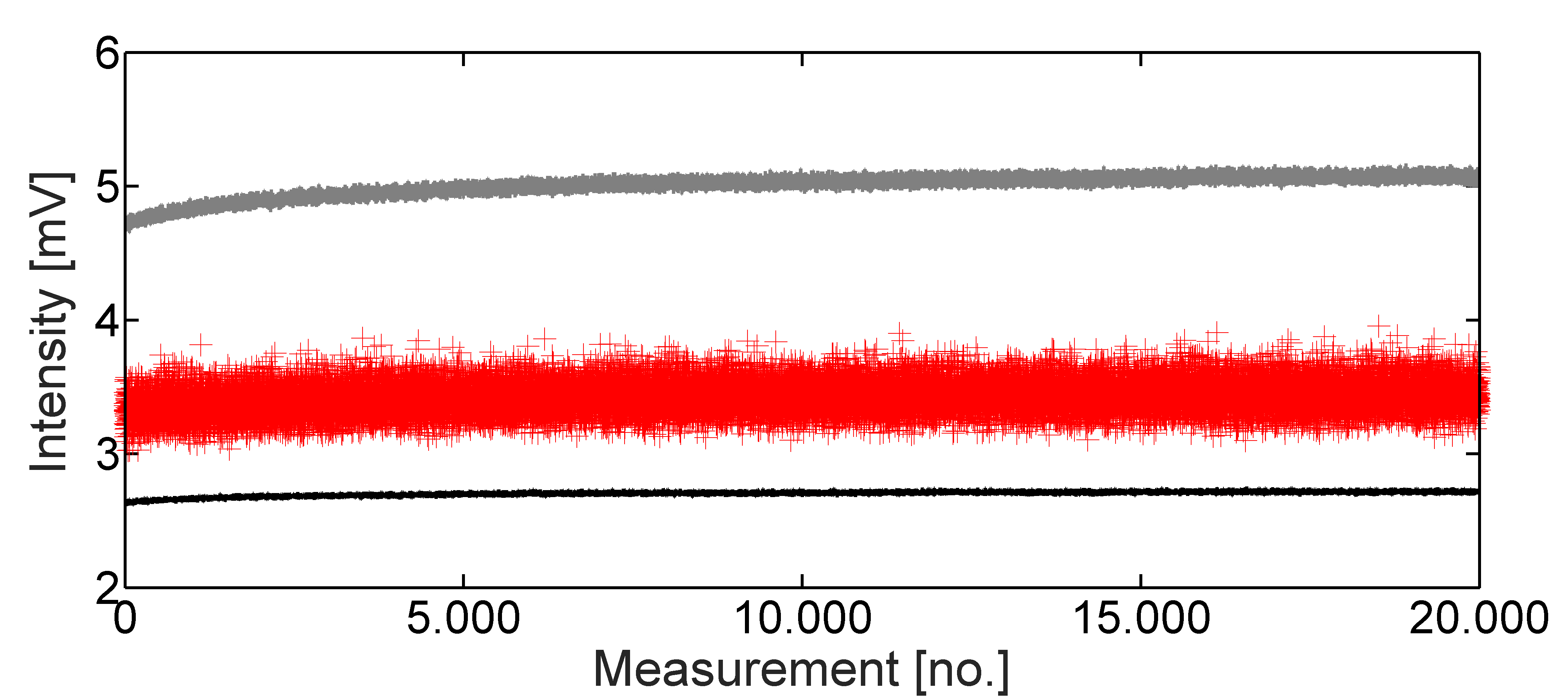}
\caption{Generic measurement signal before calibration. The figure shows the measured intensities from masks commanded to the SLM. The top and bottom lines show the measured intensities for the reference masks. The line in the middle shows the measured intensities for spatially-structured masks.}
\label{fig:sequence}
\end{figure}

A schematic representation of the experimental setup is shown in Fig. \ref{fig:setup}. The scene is illuminated by a \SI{0.35}{\tera\hertz} TX, which is a multiplier-based source emitting $\approx$\SI{7}{\milli\watt}. Through its diagonal horn antenna the TX emits an  almost Gaussian-shaped beam. The beam is then transmitted through the OS and a mapping lens made from TPX$^\copyright$. This TPX$^\copyright$ lens has a focal length of \SI{21}{\centi\meter} and a diameter of \SI{10}{\centi\meter}. It maps the THz-masks from the surface of the OS to the scene and is placed $\approx$\SI{35}{\centi\meter} behind the OS. The illuminated area measured with a Golay cell is $\approx$\SI{20}{\centi\meter} in diameter at a distance of \SI{1}{\meter} from the TPX lens. At this position objects are mounted onto a metallic mirror. The radiation that is reflected or scattered from the object/mirror is collected by a Cassegrain-type telescope at \SI{1.5}{\meter} distance from the object/mirror. The telescope has a \SI{76}{\centi\meter} diameter primary mirror and a secondary mirror with \SI{18}{\centi\meter} diameter. The field-of-view (FoV) of the telescope is matched to the \SI{20}{\centi\meter} diameter illuminated area but due to diffraction not the entire illuminated area can be used in the experiments. Finally, the radiation is detected with an InSb detector, which has a \SI{1}{\mega\hertz} bandwidth. To implement the optical modulation approach a THz-SLM is integrated in this setup.
It consists of a halogen lamp, a SLM for visible light (VIS-SLM - DMD 0.7 XGA from Texas Instruments), a Germanium (Ge) wafer, the optical switch and three VIS-lenses. The VIS-SLM has 1024 x 768 micro mirrors with a chip diagonal of \SI{17.8}{\milli\meter}. The lenses image the light from the halogen lamp onto the VIS-SLM and from there onto the Ge wafer. A dichroic mirror made from a thin indium-tin-oxide (ITO) coated glass transmits the VIS light onto the Ge wafer. At the same time it reflects the THz radiation from the TX. The Ge wafer is cut from a single-crystalline Ge rod, which has \SIrange{30}{40}{\ohm\centi\meter} resistivity and a residual n-doping of $\approx$\SI{e10}{\centi\meter^{-3}}. The wafer is \SI{5}{\centi\meter} in diameter and \SI{2}{\milli\meter} thick. The Ge wafer defines the output aperture of the THz-SLM. It translates the VIS light pattern (VIS-mask) into a THz intensity pattern (THz mask). The THz-masks are formed at the surface of the Ge wafer where the VIS light is absorbed and free electrons are generated. These areas become less transmissive for the THz radiation while those areas that are not illuminated are unaffected in their transmissibility for the radiation from the TX. Test objects were mounted  $\approx$\SI{1.3}{\meter} away from the Ge wafer onto a metallic mirror. The radiation from the TX is mechanically chopped at a frequency of \SI{1.3}{\kilo\hertz} and the signal from the InSb detector is measured with a lock-in amplifier (LIA) taking the \SI{1.3}{\kilo\hertz} as reference. 
\vspace{3pt}\\ \indent 
In order to acquire images with this setup, multiple VIS-masks have to be commanded to the SLM. This VIS-mask stream is subdivided in cycles. Each cycle consists of four \SI{3}{\milli\second}-long measurements with specific VIS-masks commanded to the SLM: 1) totally black VIS-mask, 2) spatially-structured VIS-mask, 3) totally white VIS-mask, 4) spatially-structured VIS-mask. Each time a different spatially-structured VIS-mask (pseudo-random Bernoulli mask) is commanded. The measurements with a totally black respective white VIS-mask correspond to minimum respective (no. 1) maximum THz transmission (no. 3). These measurements are used for calibration (power drift correction). The calibrated signal is determined according to a calibration procedure described in the appendix. Each image acquisition is comprised of several hundred of the measurement cycles. Aside from the calibration they also allow the derivation of an estimate for the total depth of modulation, which was about 45\% (see Fig. \ref{fig:sequence} for an example measurement). 
The plot in Fig. \ref{fig:sequence} exemplifies the measurement signal before calibration. The signal detected by the InSb detector is shown as a function of the measured number of spatially-structured masks. Each measurement corresponds to one specific THz-mask. The top  curve at about \SI{5}{\milli\volt}  corresponds to a fully transmissive mask, i.e. no VIS light shines on the Ge wafer. The line around \SI{2.5}{\milli\volt} corresponds to maximum blocking of THz radiation, i.e. the whole area of the Ge wafer is illuminated by VIS light. Each data point of the blue curve corresponds to one particular, spatially-structured THz-mask. It is important to note that the variation of the blue curve is due to the different spatially-structured THz-masks and their correlation with the scene and not due to noise. 
\vspace{3pt}\\ \indent 
The reference masks used in the measurement carry no spatial information and are as explained used for calibration purposes. The effect of the calibration procedure is exemplified in Fig. \ref{fig:calibrationExampleMain} and the impact on the acquired images is disscussed in the appendix.
\begin{figure}[H]
\centering
\includegraphics[width=\linewidth]{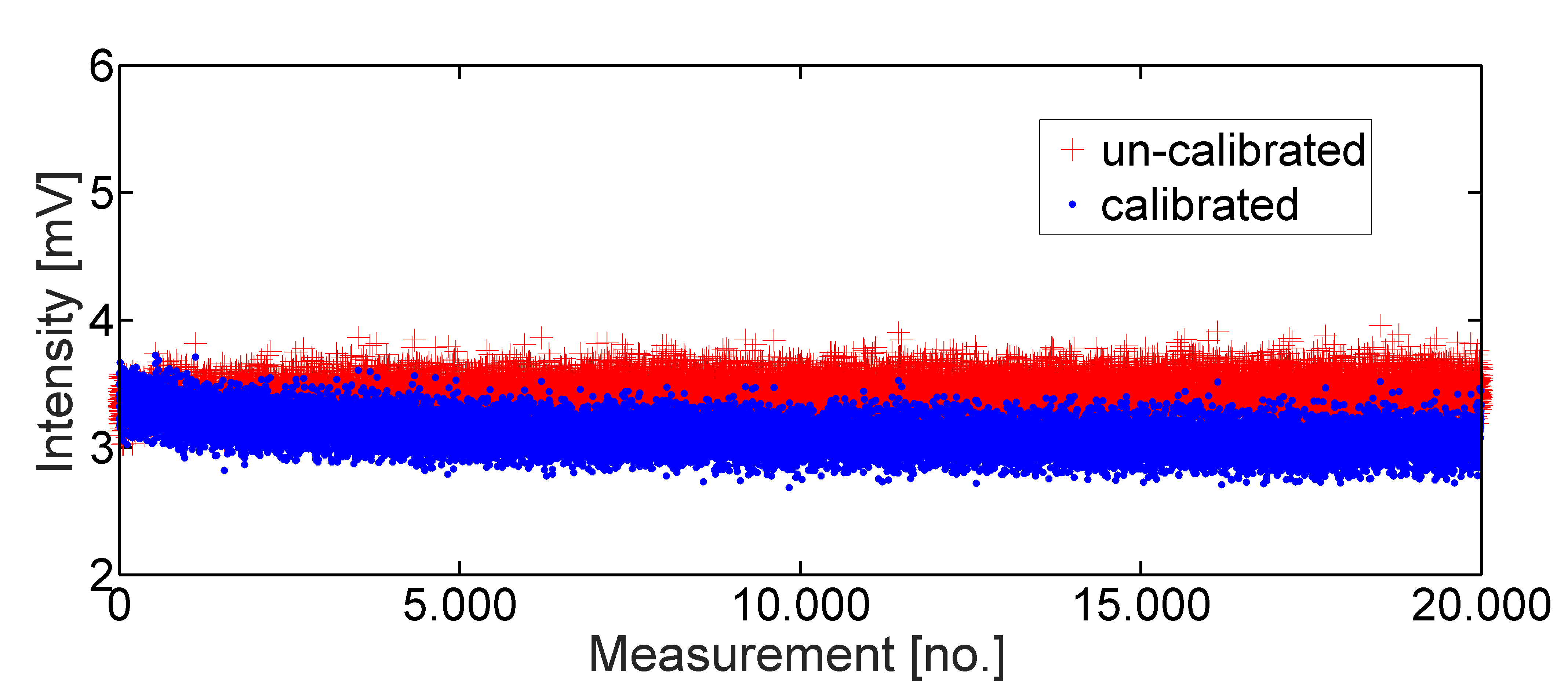}
\caption{Measurement signal of spatially-structured massks before and after calibration.}
\label{fig:calibrationExampleMain}
\end{figure}
It can be seen in Fig. \ref{fig:calibrationExampleMain} that the calibration procedure modifies the measured intensity values for all spatially-structured masks and that the changes in intensity can be as large as several percent of the measured value. The calibration procedure was applied to all measurements presented in the main text. It was found that the calibration modifies the resulting images but does not change the fundamental nature of the reconstructed images. Further details of the calibration procedure are discussed in the appendix.   
%
\subsection{Compressed Sensing protocol}\label{sec:CS}
Compressed sensing is a novel mathematical research area that was introduced in 2006 \cite{Candes.2006},\cite{Donoho.2006}. It predicts that many high-dimensional, real-world signals  (compressible/sparse signals) can be recovered from incomplete measurements. Applied in an imaging context Compressed Sensing is one application example of computational/mathematical imaging with all potential advantages of a computational imaging approach \cite{Hunt.2013} including the potential for substantial undersampling \cite{Lustig.2008},\cite{Donoho.2009}.
\vspace{3pt}\\ 
\indent The prerequisites for achieving the advantages stated in the references, optimally in terms of sampling rate, are the precise knowledge of the intensity patterns (THz-masks) used to illuminate the scene and a certain randomness in these masks. The randomness allows to acquire a large spatial frequency content with each measured spatially-structured mask. This in turn allows for the undersampling capability of Compressed Sensing and also for its non-adaptivity properties \cite{Iwen.06.11.201109.11.2011},\cite{Do.2012}. The randomness has another advantage that is very useful here. In cases where the THz-masks used to illuminate the scene are degraded somehow, meaning that the THz-masks are altered in the imaging process the randomness allows for an averaging effect that makes it possible to reconstruct a coarse image. The mask degradation from an initial state where the masks are completely known to a degraded state that is relevant for the image reconstruction presents a special challenge (see Section \ref{sec:challenge}) in a stand-off modality.  
\vspace{3pt}\\ \indent 
The Compressed Sensing protocol that was used in the experiments, is essentially an implementation example of an active THz single detector pixel camera \cite{Duarte.2008} projecting pseudo-random binary masks combined with an $\ell_1$/$\ell_2$-regularized reconstruction approach. Therefore, the Compressed Sensing protocol combines the hardware- and software level namely the randomness in the masks and special algorithms to solve the inverse problem. The decoding/reconstruction process can be achieved by several techniques as long as the masks used in the acquisition process are known (otherwise the image reconstruction will fail). Depending on the chosen algorithm, the required spatial resolution, the undersampling ratio, degradation of the masks, or imaging model errors the results of the decoding process are of varying quality. For the results presented here, we have chosen a non-negative least squares (NNLS) algorithm as decoding procedure. This choice is based on the robustness and sparsity promoting properties of NNLS. It is a $\ell_2$-minimization approach with the additional constraint of non-negativity. This minimization approach can be rewritten as a $\ell_1$-minimization when pseudo-random binary masks are used \cite{Kueng.2017}. Therefore, this approach falls under the subject of Compressed Sensing algorithms.
\vspace{3pt}\\ \indent 
The Compressed Sensing protocol is completed by the use of a linear imaging model $ y = A\cdot x + n$. Here, $y$ is the vector of the measured intensity values, $x$ is the THz representation of the scene to be reconstructed and $n$ is an additive noise component. The measurement matrix $A$ contains in the rows the masks commanded to the SLM. Due to the large distance from the SLM to the scene it is highly unlikely that the protocol will facilitate undersampling without additional knowledge of the mask degradation. In summary, the most prominent features of our Compressed Sensing (CS) protocol, namely the ability to measure large spatial frequency content with every spatially-structured mask measurement (pseudo-random masks) and the advantage that with every structured mask measurement the signal-to-noise ratio (SNR) of the acquired images increases, are crucial. These features enable a successful image acquisition at stand-off distances, which will be exemplified in the next section.
\section{Results} 
\begin{figure}[H]
\centering
\includegraphics[width=\linewidth]{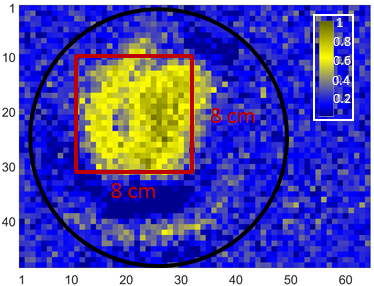}
\caption{Reconstructed beam of the TX reflected from a metallic mirror in the scene. The red frame indicates the region which is used for the imaging experiments and the black circle indicates the region of non-zero modulation.} 
\label{fig:fov}
\end{figure}

\begin{figure}[H]
\centering
\includegraphics[width=\linewidth]{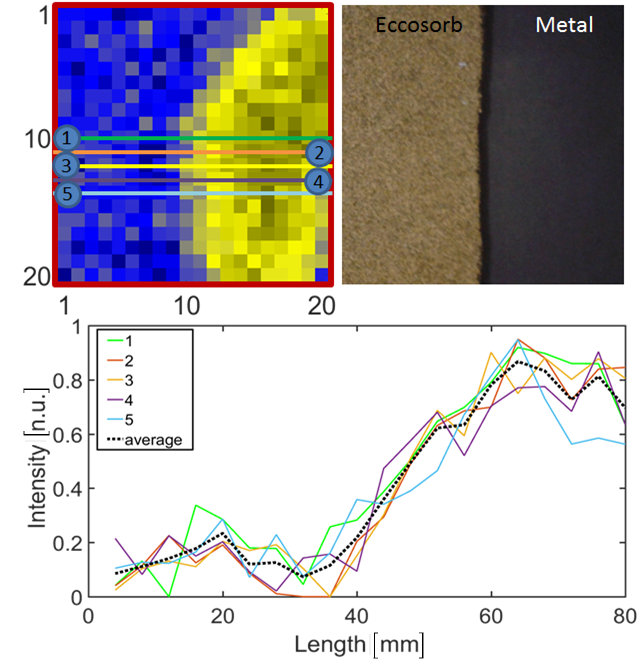}
\caption{Edge analysis of an Eccosorb$^\copyright$ test target that blocks half of the FoV (cf. red frame in Fig. \ref{fig:fov}). The figure shows cuts across the Eccosorb$^\copyright$ edge at the center of the FoV. From the average of these cuts an edge fidelity  of \SI{2}{\centi\meter} - \SI{2.5}{\centi\meter} can be derived if  the intensity increase from 10\% to 90\% is used as criterion.}
\label{fig:edge}
\end{figure}

\begin{figure}[H]
\centering
\includegraphics[width=\linewidth]{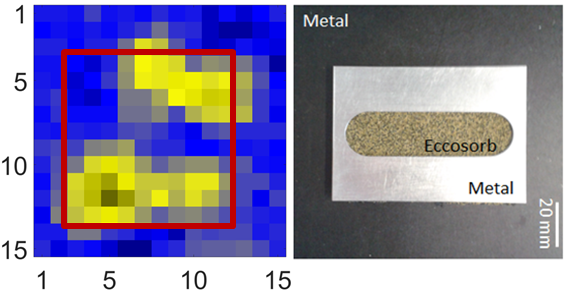}
\caption{Reconstructions of a full reflection mode metallic test target. Since the target has a size of 60x\SI{80}{\milli\meter}$^2$ a slightly larger portion of the scene is shown (notice again the red frame and compare with Fig. \ref{fig:fov}).}
\label{fig:reflectiontarget}
\end{figure}

\begin{figure}[H]
\centering
\includegraphics[width=\linewidth]{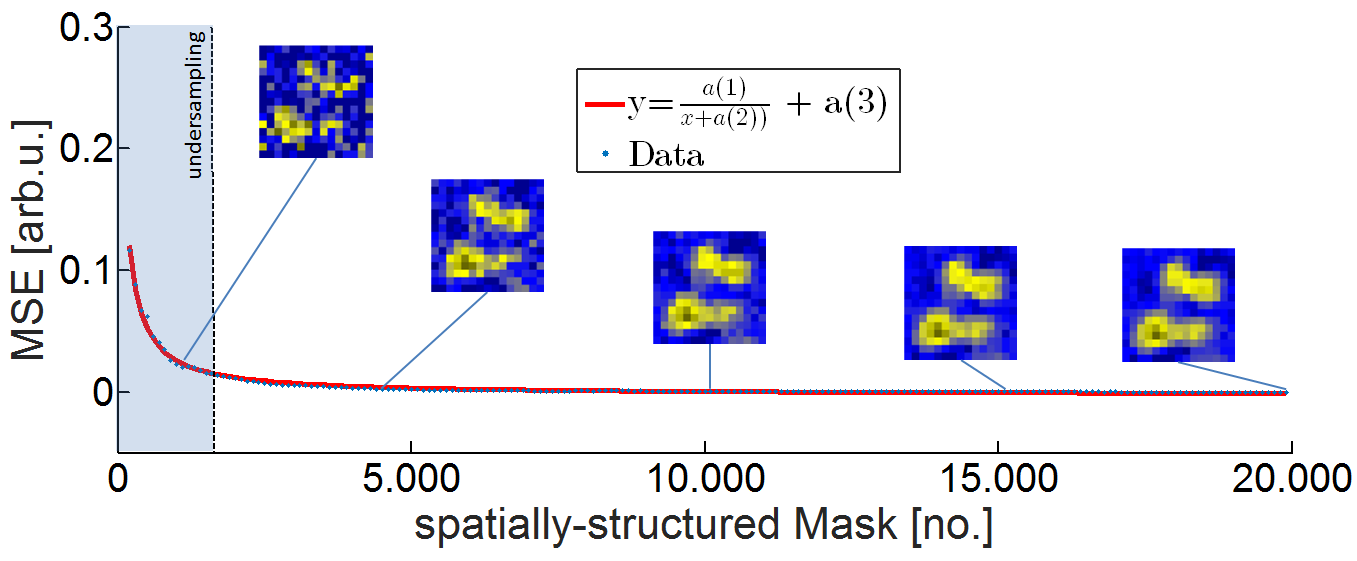}
\caption{Analysis of mean squared error development with varying number of measurements (number of spatially-structured masks) of the target shown in Fig. \ref{fig:reflectiontarget} demonstrating the influence of various effects (averaging) on the reconstruction quality. The dotted line in the figure marks the threshold of the number of pixels (Shannon case).} 
\label{fig:mse}
\end{figure}

With the setup shown in Fig. \ref{fig:setup} we first analyzed the achievable field-of-view (FoV) by imaging the reflection of the TX beam from a metallic mirror. The reconstructed image of this measurement is shown in Fig. \ref{fig:fov}. It consists of 64 x 48 pixels. At the DMD this corresponds to 16 x 16 mirrors per pixel (block size) which in turn corresponds to a square of 1 x \SI{1}{\milli\meter}$^2$ at the OS. We have chosen this size as a compromise between diffraction at the openings of the THz-masks and the number of pixels in the THz-image. It should be noted that at the scene each image pixel has a size of approximately 4 x \SI{4}{\milli\meter}$^2$. 
As can be seen in Fig. \ref{fig:fov}, the usable FoV is limited by a diffraction artifact, namely the dark, ring-shaped area around the central maximum. We have analyzed the optics using ZEMAX simulation software and it turns out that the ring structure is mainly caused by diffraction at the secondary mirror of the Cassegrain telescope. The size of the usable FoV is, therefore, limited to an area of approx. 8 x \SI{8}{\centi\meter}$^2$ inside the diffraction ring (red frame in Fig. \ref{fig:fov}). This area consists of 20 x 20 pixels. Throughout the rest of this paper we will show results, which have been measured within this FoV (red frame). However, it should be also noted that the reconstruction algorithm is always applied to the full image with 64 x 48 pixels. Additionally, only within the FoV the linear CS model discussed in the previous section is a good approximation and can be reliably used for image reconstruction. 
Fig. \ref{fig:fov}. shows the entire FoV of our setup. The black circle in the image indicates the region of non-zero modulation enabled by the OS. The reconstructed pixels that are outside of this circle are unreliable and are image noise caused by the reconstruction process. Mainly, the noise is caused due to the fact that the  OS has a circular shape and the DMD as described has a rectangular shape. The mapping of the VIS-masks was adjusted in such a way that the vertical axis of the DMD was mapped on the diameter of the OS. That results in regions in the reconstructed images were the depth of modulation is zero. These regions are used to provide an estimate for the measurement noise respective the achievable contrast in the reconstructed images.   
\vspace{3pt}\\ \indent 
The next step of the experiments was the determination of edge fidelity using a so-called half block target with a single edge at the center of the FoV (see Fig. \ref{fig:edge}). As mentioned, we have determined the edge fidelity of our imaging system by covering half of the FoV with Eccosorb$^\copyright$  (see Fig. \ref{fig:edge}). This way a single edge at the center of the FoV is created. A quantitative analysis of the edge fidelity is also shown in the figure, in the plot on the bottom. The curves display the intensity along horizontal cuts through the image. The intensity grows from approximately 10\% to 90\% within about \SI{25}{\milli\meter} which corresponds to approximately six image pixels. This edge fidelity is a first indication for the achievable spatial resolution of the setup when VIS-masks with block size 16 x 16 are used. 
\begin{figure}[H]
\centering
\includegraphics[width=\linewidth]{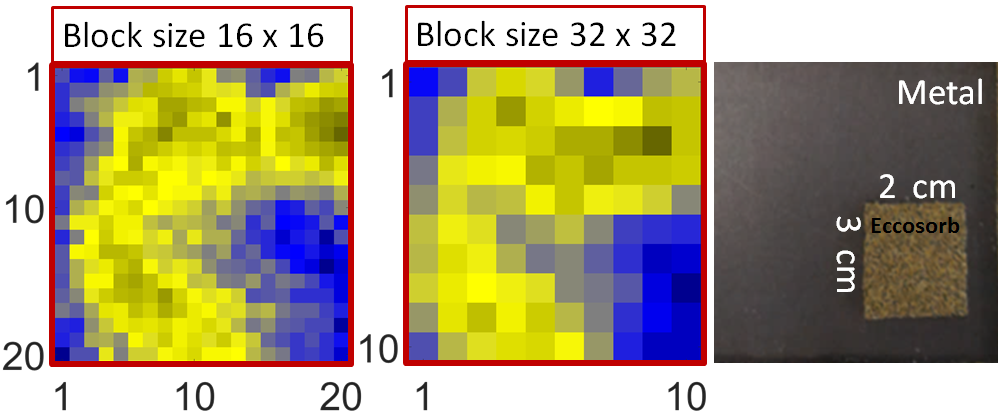}
\caption{Small Eccosorb\copyright\ test target with horizontal and vertical edges inside the FoV. Shown are on the left-hand side a reconstruction with block size 16 x 16 and in the middle with block size 32 x 32. A photograph of the target is shown on the right-hand side. }
\label{fig:small}
\end{figure}
In order to get an improved measure we imaged another test target much smaller than the half block. This test target is also made from Eccosorb$^\copyright$ but covers only the bottom right corner of the FoV (see Fig. \ref{fig:small}). 
The figure shows two reconstructions with different block size of the commanded masks. A block size of 16x16 (left) and 32x32 (middle) were used resulting in a reconstruction with 400 pixels and 100 pixels, respectively. For direct comparison the target is shown on the right-hand side. In both THz images the horizontal as well as the vertical edge are visible. However, in the image which was acquired with a block size of 16x16 the reconstruction seems to be somewhat worse, possibly because the SNR of each pixel is reduced due to the four times smaller pixel size and the correspondingly lower power per pixel. This is more pronounced towards the border of the FoV where the power from the TX is lower. In order to evaluate the full reflection mode imaging capabilities in more detail, a metallic slit target was imaged as final test target (see Fig. \ref{fig:reflectiontarget}).
It consists of a 60x\SI{80}{\milli\meter}$^2$ aluminium plate with a \SI{20}{\milli\meter} wide and \SI{70}{\milli\meter} long slit. This target was placed on top of an Eccosorb$^{\copyright}$ absorber and was then mounted at the metallic mirror in the scene (see photograph on the right-hand side in Fig. \ref{fig:reflectiontarget}). For an improved SNR, the THz image was taken with a block size of 32x32. Accordingly, in the scene each pixel corresponds to 8 x \SI{8}{\milli\meter}$^2$. The metal above and below the slit is clearly visible as two bright, longitudinal reflections while the vertical borders, which are only \SI{5}{\milli\meter} wide, are not resolved. It should be noted that the image like all other presented results have been reconstructed using 20,000 binary, pseudo-random, spatially-structured masks, i.e. several times more masks than image pixels. It is assumed that this large amount of oversampling was necessary due to a required averaging effect to increase the measurement SNR. Several effects can be responsible for the decrease in SNR e.g. the fast measurement speed, small integration time as well as the mask degradation effect due to the stand-off distance that was already mentioned before. 
\vspace{3pt}\\ \indent 
As an additional analysis of the averaging effect, we investigated the reconstruction quality with varying number of masks. For this analysis we reconstructed the image shown in Fig. \ref{fig:reflectiontarget} using a varying number of masks, starting from 200 spatially-structured masks up to 20,000 and measured the reconstruction quality using the mean squared error (MSE) of the reconstructed images as a quantitative quality indicator. The analysis is shown in Fig. \ref{fig:mse}. It is visible that with an increasing number of spatially-structured masks the image quality improves significantly. With only 500  spatially-structured masks the two reflections from the horizontal bars are already recognizable (undersampling) albeit not completely. The image quality continues to improve but apparently no further details become visible. 
While the numeric details of the analysis are given in the appendix a closer look at Fig. \ref{fig:mse} already reveals additional details about the degradation. As stated, in the first reconstructed image shown, the number of spatially-structured masks is 500, which is less than the number of image pixels (undersampling regime). But, already the two metallic bars of the test target (see Fig. \ref{fig:reflectiontarget}) are recognizable albeit with poor image quality. With increasing number of masks the image quality steadily improves while the MSE decreases by almost two orders of magnitude. As is known, in a fully ideal setting, without any degradation effects, the decrease in MSE value should follow a $y=1/x$ characteristic. 
\vspace{3pt}\\ \indent 
However, the fit that is also shown in the figure, demonstrates that the decrease is modified by several constants a(1) - a(3) but that the decrease in MSE follows the $y=1/x$ characteristic with only slight modifications. In other words, with increasing number of measured masks the deviation from the ground truth approaches zero. As ground truth the reconstruction using 20.000 spatially-structured masks was used for calculating MSEs. This exemplifies the averaging effect due to an increase in SNR.       
\vspace{3pt}\\ \indent 
%
\section{Discussion} 
In the previous section we have shown that the optical modulation technique  can be applied to stand-off THz imaging. The technique is based on a structured illumination of the scene in combination with a single pixel detector \cite{Phillips.2017}. For a structured illumination approach the fidelity of the THz-masks at the scene determines almost all relevant image parameters e.g. the achievable spatial resolution as well as the SNR of the image. However, the usable FoV of the measurements was not limited by the illumination. It was limited by diffraction at the Cassegrain telescope and its secondary mirror to a size of \SI{8}{\centi\meter} x \SI{8}{\centi\meter}. Another source of diffraction is the TPX$^\copyright$ mapping lens, which has a diameter of only \SI{10}{\centi\meter}. The diffraction from the mapping lens does not only potentially limit the FoV, more importantly it is a potentially relevant cause for the degradation of the THz-masks. The degradation occurs along the distance from the illuminated surface of the OS to the scene and it stems mainly from the divergence of the THz beam and from diffraction effects. Additionally, the coherent nature of the THz radiation leads to interference that can distort the THz-masks at the scene. The effect of mask degradation limited the achievable spatial resolution in the presented images and needs to be overcome in order to successfully implement a security imaging application using the optical modulation approach described here.   
%
\section{Conclusion}
In conclusion, the results demonstrate the applicability of our optical modulation approach with structured illumination for stand-off imaging but they also show the current limitations of the approach.  
\vspace{3pt}\\ \indent 
The acquired signals of the image acquisition process are encoded and in order to produce a human readable image representation of the scene, the acquired signals need to be decoded using a reconstruction procedure/decoding algorithm. The complexity of the necessary decoding procedure is thereby directly connected to the nature of the masks respective the acquisition process itself. This in turn provides direct control of several image parameters. For example, the direct control of the achievable spatial resolution, the SNR, etc. simply by varying the block size of the masks. This makes the dynamic aperture imaging approach especially suited for applications in the THz region of the electromagnetic spectrum, especially when used in a reflection geometry. In order to overcome the limitations by mask degradation a better understanding of the masks is required. For example taking diffraction explicitly into account e.g. in a non-linear imaging model (phase retrieval). When directly implemented in the reconstruction process this will certainly improve the image quality. This approach falls under the subject of lens-less THz imaging, which is a direction that would overcome many of the present limitations. With such an improvement, applications like stand-off security imaging could profit greatly from the  advantages of THz dynamic aperture imaging presented here. 
In summary the versatility of an optical modulation approach combined with the ability to apply it at stand-off distances may open new directions for THz imaging.

%
\appendices
\section{Calibration Procedure}
\begin{figure}[H]
\centering
\includegraphics[width=0.9\linewidth]{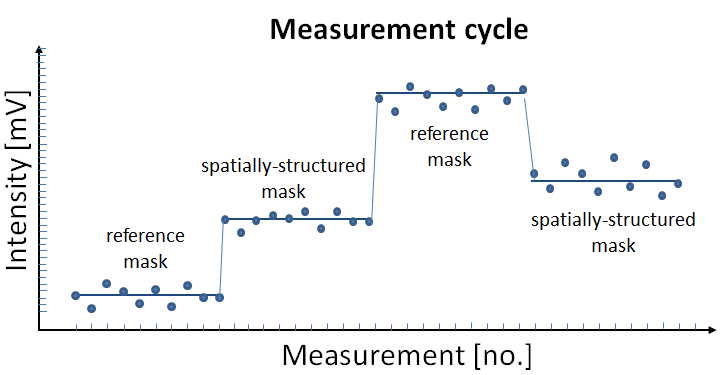}
\caption{Visualization of a measurement cycle of four masks with two spatially-structured-masks and two un-structured masks (reference). For each mask 10 intensity values are measured from which, for each mask, the median is computed. The resulting intensity value is saved and used for reconstruction (y).}
\label{fig:cycle}
\end{figure}

\begin{figure}[H]
\centering
\includegraphics[width=\linewidth]{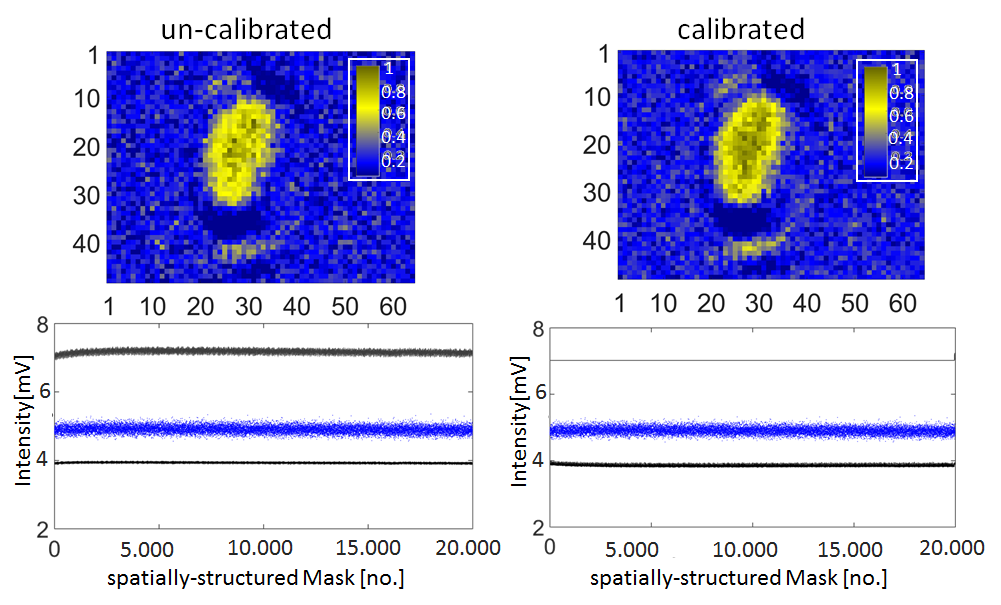}
\caption{Example measurement visualizing the impact of the calibration procedure. The figure shows in the top row the images reconstructed from calibrated/un-calibrated measurement signals. The measured signals itself are shown in the bottom row.}
\label{fig:example}
\end{figure}

The acquired signals are subject to power drifts and instabilities of the used halogen lamp. In order to account for these instabilities a calibration procedure is used. This procedure has no need for an additional reference detector, instead it uses spatially un-structured masks, which are included in the measurement process solely for the purpose of calibration. The measurement of these masks produces no spatial information, instead they produce correction factors that are used to calibrate spatially-structured mask measurements. In regard to an image measurement the un-structured masks constitute measurement overhead. In order to keep this overhead to a minimum, an entire image measurement is divided into cycles and each cycle is comprised of four masks, two spatially-structured masks and two un-structured masks. Of the un-structured masks one is totally black and one is totally white. This corresponds to no light on the optical switch and the optical switch completely illuminated. The entire image acquisition is comprised of several hundred 
of the described cycles. One cycle is exemplified in Fig \ref{fig:cycle}.
The specific number of the necessary cycles depends on several factors, for example the measurement SNR, the VIS-mask fidelity on the OS and the THz-mask fidelity in the scene. The influence of some of these effects on the image quality can be accounted for and some can even be overcome. We are concerned here with a calibration procedure that is used to overcome the effect of power drift. As an example of the calibration procedure Fig. \ref{fig:example} visualizes its influence.  
%
The effects of the calibration procedure are apparent. It changes the measured signals and the resulting images. The visible effects in the images appear to be small but the difference in the pixel values is approximately 25\%. However, the impact of the calibration becomes even more pronounced when higher spatial resolution is aspired (smaller block size) or when the imaging process is plagued by severe power drifts. This was not the case in the image shown in Fig. \ref{fig:example} and the image was measured with a block size of only 16 x 16 SLM mirrors per block. The key point of the calibration procedure is the use of un-structured masks in order to derive correction factors for the intensity measured when spatially-structured masks are commanded to the SLM. Essentially, the correction factors are derived under the requirement that the intensity measured for the unstructured reference mask (white mask) remains the same during the time of an entire image acquisition. The calibration algorithm derived from this requirement is given in Algorithm \ref{algo:calib}.  
\begin{algorithm}
  \caption{Calibration Procedure}
  \label{algo:calib}
  \begin{algorithmic}[1]
    \State for i=1:size(data,2)/2 \label{op0}
    \State y(i)=y(i).*white(1)/white(i);
    \State end
  \end{algorithmic}
\end{algorithm}
As can be seen, the algorithm adjusts all measured intensities of the spatially-structured masks. These intensities are saved in the variable y. It uses the measurement of the first white mask as reference and adjusts the intensities in the variable y accordingly. After the application to the spatially-structured masks the same procedure is applied to the reference masks themselves. This is done for validation as well as to determine the residual variations that can not be calibrated by the procedure. Please keep in mind that only half of the measured intensities are spatially-structured measurements. The second reference mask, which blocks no light on the OS, is not used for calibration but is instead used to track the changing depth of modulation and the level of the residual variations. The measured intensity values when these masks are commanded to the SLM are included into the measurement matrix $\Phi$ to improve reconstruction quality. %
As mentioned, the greatest advantage of the calibration procedure is that a reference detector is not necessary. This is especially beneficial since the use of a reference detector is quite challenging in a single detector pixel setup. On the other hand, a substantial measurement overhead is introduced into the imaging process. However, a drawback of the calibration procedure is that it only accounts for power drifts. 
\vspace{3pt}\\ \indent
Details regarding the mean squared error analysis given for Fig. \ref{fig:mse} of the main part of this document are discussed in the final section of this appendix and additional details about the nature of the necessary oversampling are given.    
\section{Mean Squared Error Analysis}
As is known, the MSE is a quality indicator that uses the average of the sum of quadratic deviations from a ground truth to measure image quality (see (\ref{eqn:mse})).
\begin{equation}\label{eqn:mse}
MSE= \frac{1}{M}\cdot\sum\limits_{i=1}^{M}\left(P_i-Ref_i\right)^2
\end{equation}
In (\ref{eqn:mse}) M denotes the number of pixels in an image and $P_i$ is the numeric value of pixel i. Lastly, the value of the ith pixel in the reference image (ground truth) is denoted $Ref_i$. 
\vspace{3pt}\\ \indent
The MSE is one quantitative value to describe the deviation of the entire image in question from the ground truth. When used for a single image the absolute value of the MSE is not really indicative, regarding the deviation from the ground truth. Its true merits come into play when a series of images needs to be evaluated. Due to this property in combination with its simplicity the MSE was chosen as image quality indicator. As an example the performance of MSE is visualized in an ideal single detector pixel imaging setting (simulation) in Fig. \ref{fig:mse}
Ideality in this context means that the masks commanded to the SLM are equal to the VIS-masks at the OS and the THz-masks at the scene. Additionally, a depth of modulation of 100\% is used. The figure shows a few image reconstructions of simulated mask measurements, for a noise-free (top row) and a noisy case (bottom row - SNR $\approx$14.7), reconstructed using the NNLS algorithm. The reconstructed images can be used to visually evaluate the image quality and to get a feeling for their respective MSE value. Also shown is a characteristic for the MSE with increasing number of measured, spatially-structured masks. A  best fit for the MSE characteristic shows that only in the ideal, noisy setting the MSE improves with the expected y=1/x characteristic when the number of measured, spatially-structured masks is increased. For the simulation we used a scaled down photo as ground truth (reference) to compare the reconstructions to. It should be noted that the ideal noise-free setting is free of additional noise but the visible averaging comes from remaining noise introduced on the software level by the reconstruction/decoding process. 
%
\begin{figure}[H]
\centering
\includegraphics[width=0.95\linewidth]{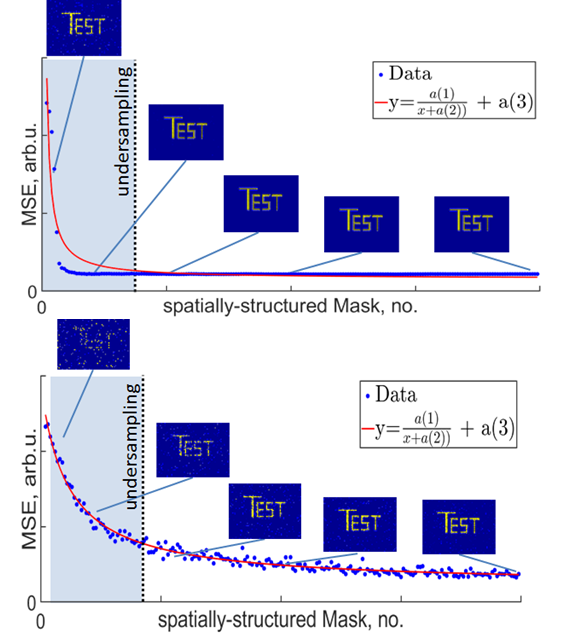}
\caption{MSE characteristic for an ideal single-pixel imaging setting. Top row shows the characteristic without added noise and the bottom row with added Gaussian noise (SNR$\approx$14.7).}
\label{fig:mseSM}
\end{figure}
%
The analysis shown in the main part of this document in Fig. \ref{fig:mse} used the same analysis procedure but it used the reconstruction with 20.000 spatially-structured masks as ground truth. Also, the analysis shown in Fig. \ref{fig:mseSM} used a block size of 16 x 16 whereas the analysis shown in Fig. \ref{fig:mse} used a block size of 32 x 32. The smaller block size was used here to adequately visualize the influence of several degradation effects. When this analysis is compared with the analysis in the main part of the document, several observations can be made. Firstly, the best fit in Fig. \ref{fig:mse} follows a characteristic that is modified by several constants according to y = a$_1$/(x+a$_2$) + a$_3$. The least-squares fit gives a$_1$=29.7, a$_2$=41.6 and a$_3$=-0.0025 with an optimality tolerance of less than 1e-06. The 1/x characteristic documents the averaging that is necessary due to several non-idealities (noise, imaging model errors, algorithmic effects and THz mask degradation, etc.). The constants in the fit only document the existence of non-idealities but do not indicate which effects are present. However, when compared with the analysis of the ideal setting the large amount of necessary oversampling in combination with the good SNR suggests THz mask degradation as possible explanation. However, to proof without a doubt that really mask degradation is responsible is not possible at the current technology level.

\section*{Acknowledgment}
The authors are very grateful to Dr. Nikolay V. Abrosimov from the Leibniz Institute for Crystal Growth in Berlin for supplying the Germanium crystal from which the OS was cut that was used in the experiments. 
This work was funded thanks to the German Research Foundation (DFG) under the grants HU848/7 and JU2795/3.
%

\vspace{0pt}
\begin{IEEEbiography}[{\includegraphics[width=1in,height=1.25in,clip,keepaspectratio]{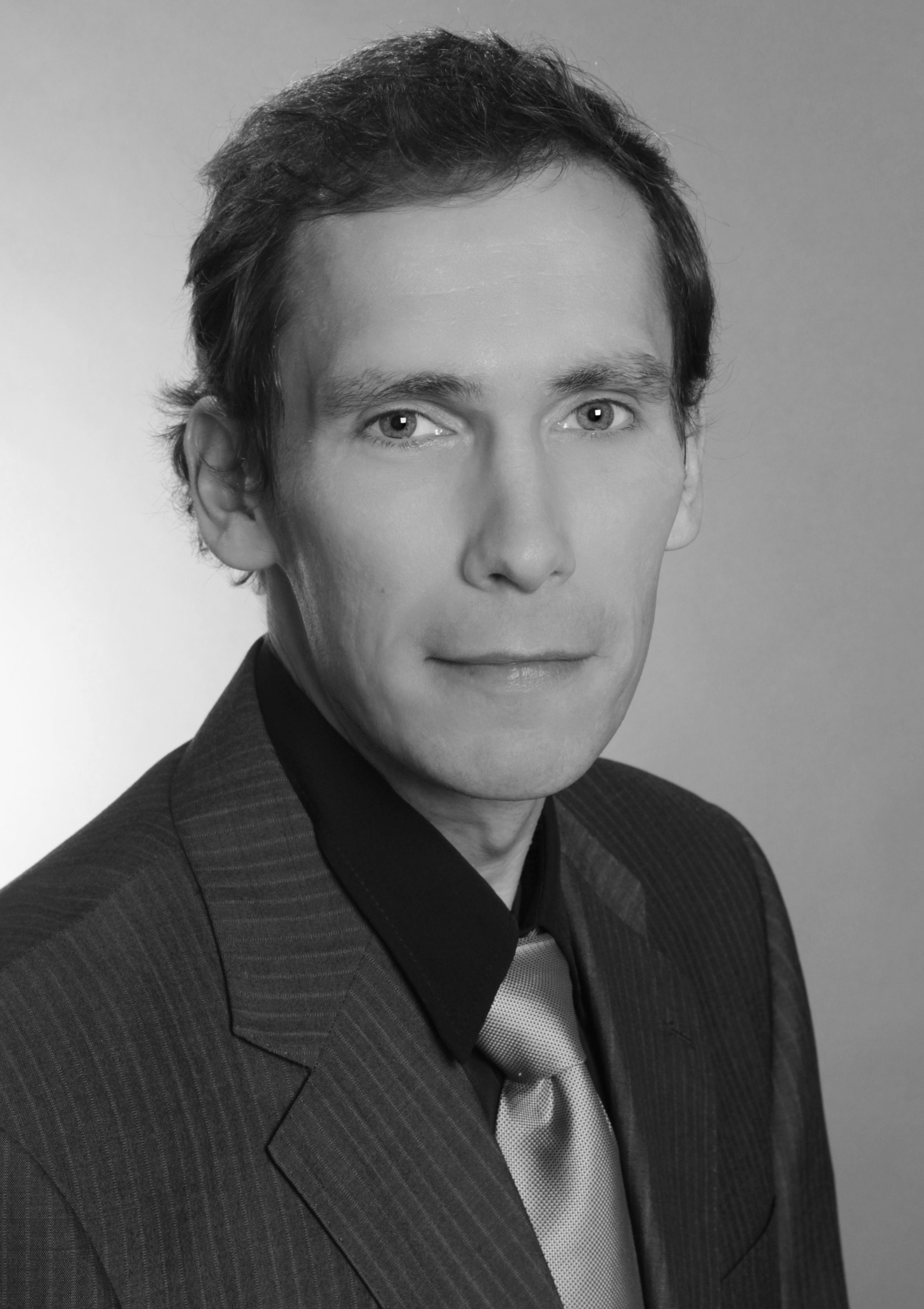}}]{Sven Augustin}
is a member of the German Physical Society. He received his diploma in physics from free university Berlin and the Dr. rer. nat. from Technische Universität Berlin in 2015 for his work on terahertz security imagers. He has a background in THz image processing and works on the subject of THz single-pixel imaging respective THz dynamic aperture imaging since 2013. He is especially interested in applying THz dynamic aperture imaging for security tasks and has until recently worked in the DFG priority program "Compressed Sensing in Information Processing (CoSIP)" and is now using the advances made while working on the project to develop THz security imaging based on a Compressed Sensing measurement scheme.    
\end{IEEEbiography}
\vspace{-200pt}
\begin{IEEEbiography}[{\includegraphics[width=1in,height=1.25in,clip,keepaspectratio]{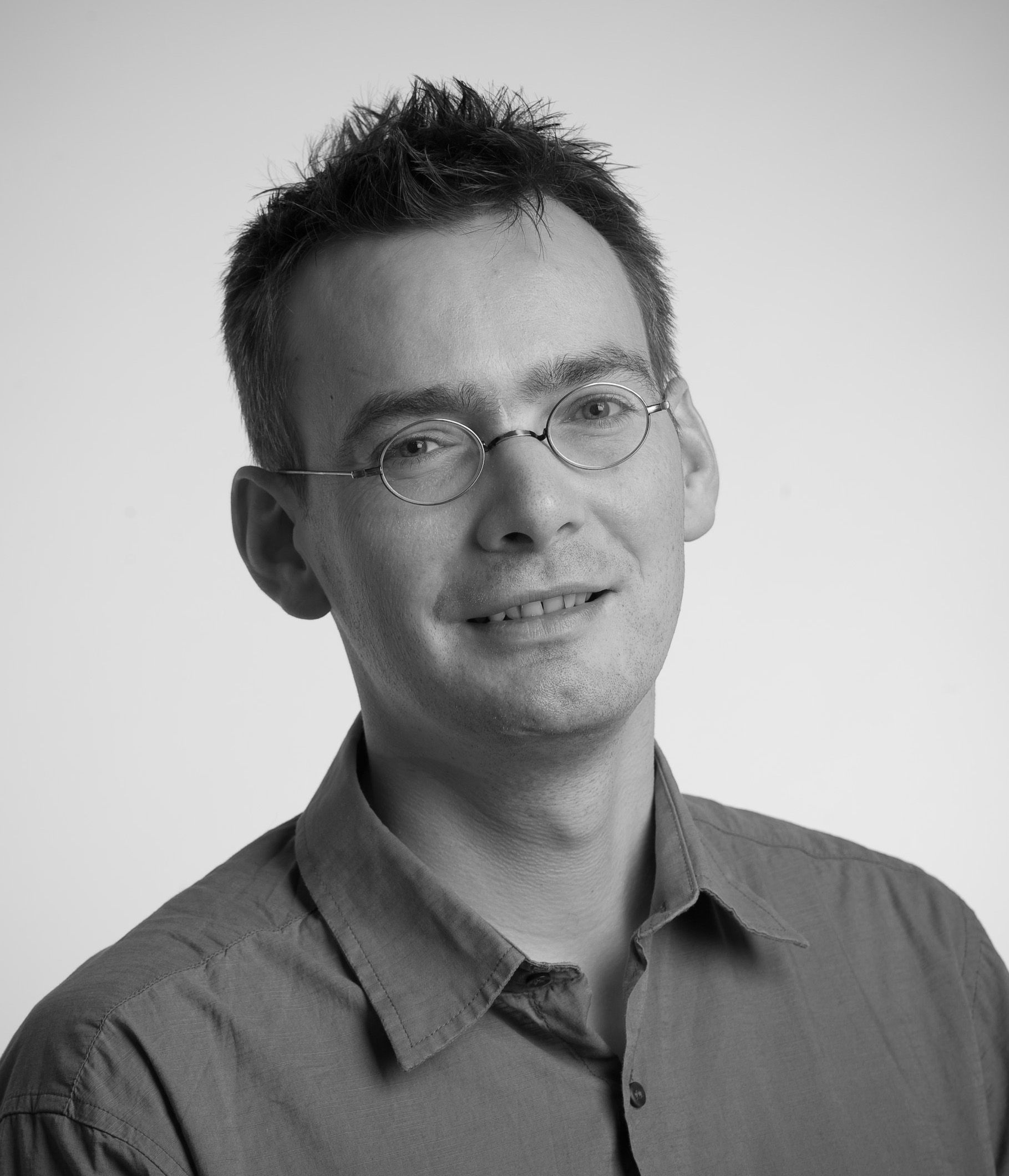}}]{Peter Jung}
 (Member IEEE, Member VDE/ITG) received
  the Dipl.-Phys. in high energy physics in 2000 from Humboldt Universität zu Berlin, Germany, in cooperation with DESY Hamburg. Since
  2001 he has been with the Department of Broadband Mobile Communication Networks, Fraunhofer Institute for Telecommunications,
  Heinrich-Hertz-Institut (HHI) and since 2004 with Fraunhofer German-Sino Lab for Mobile Communications.  He received the
  Dr. rer. nat degree in 2007 (on Weyl--Heisenberg representations in communication theory) from the Technische Universität Berlin, Germany. P. Jung is currently working under DFG grants at the Technical University in Berlin, Germany
  in the field signal processing, information and communication theory and data science.  His current research interests are in the
  area compressed sensing, machine learning, time--frequency analysis, dimension reduction and randomized algorithms.  He is giving
  lectures in compressed sensing, estimation theory and inverse problems.
\end{IEEEbiography}
%
%
\vspace{-200pt}
\begin{IEEEbiography}[{\includegraphics[width=1in,height=1.25in,clip,keepaspectratio]{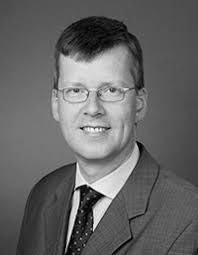}}]{Heinz-Wilhelm Hübers}
received a diploma degree and Dr. rer. nat. degree in physics in 1991 and 1994, respectively, from the Universität Bonn, Germany. From 1991 to 1994, he was with the Max-Planck-Institut für Radioastronomie in Bonn, Germany. In 1994, he joined Deutsches Zentrum für Luft- und Raumfahrt (DLR) in Berlin, Germany, where he became head of the department "Experimental Physics" in 2001. In 2009 he became professor of experimental physics at the Technische Universität Berlin, Germany. In 2015 he changed to Humboldt Universität zu Berlin and is now professor of optical systems at the Humboldt Universität zu Berlin and the director of the institute of Optical Sensor Systems at DLR Berlin. His research interests include terahertz technology in particular lasers and detectors, and the application of terahertz technology to planetary research, astronomy, earth observation, and security. He is a member of the Deutsche Physikalische Gesellschaft (German Physical Society) and the Verein Deutscher Ingenieure (Association of German Engineers). He has received the Innovation Award for Synchrotron Radiation (2003) and the Lilienthal Award (2007).
\end{IEEEbiography}

\end{document}